# Non-Equilibrium Fluidization of Dense Active Suspension

Yujiro Sugino[1], Hiroyuki Ebata[1], Yoshiyuki Sowa[2], Atsushi Ikeda[3,4], and Daisuke Mizuno[1]

1. Department of Physics, Kyushu University, 819-0395 Fukuoka, Japan.
2. Department of Frontier Bioscience, Hosei University, 184-8584 Tokyo, Japan
3. Graduate School of Arts and Science, The University of Tokyo, 153-8902 Tokyo, Japan
4. Universal Biology Institute, The University of Tokyo, 153-8902 Tokyo, Japan

22 January 2024

**Abstract**

We investigate dense suspensions of swimming bacteria prepared in a nutrient-exchange chamber. Near the pellet concentration, nonthermal fluctuations showed notable agreement between self and collective behaviors, a phenomenon not observed at equilibrium. The viscosity of active suspensions dramatically decreased compared to their inactive counterparts, where glassy features, such as non-Newtonian viscosity and dynamic heterogeneity, disappeared. Instead, the complex shear modulus showed a power-law rheology, $G^*(\omega) \propto (-i\omega)^{1/2}$, suggesting the role of bacterial activity in driving the system towards a critical jamming state.

Active matter refers to a group of interacting objects that convert ambient energy from environment into mechanical work. With continuous input of mechanical energy, active matters often exhibit collective motion and self-organize dissipation structures, which does not obey equilibrium physics [1,2]. A colloidal glass is an opposite. In the absence of nonequilibrium activities, the dynamics of colloidal objects slows down dramatically when crowded. Such inert glasses do not reach equilibrium on humanly observable time scales, which has brought up conundrums to physicists [3]. Recent theoretical studies suggested that colloidal glasses may melt at room temperature when each colloid generates active forces. A situation both dense and fluid, a conflicting requirement at equilibrium, can then be achieved. This possibility brings us to investigate the dense active matter made of glass-forming materials, referred to as "active glass".

Previous theoretical studies predicted that active forces fluidize a glass by facilitating structural relaxations [4-11]. However, opposite results have been also reported. Active forces further decelerate glassy dynamics by fostering dynamic heterogeneity and/or aging process [12-16], responsible for the glassy slow dynamics. So far, existing studies have focused on the fluctuation of particles by relying on numerical simulations while a fundamental property, i.e., rheology, has not been investigated. Elevated fluctuations in active glass does not necessarily imply its fluidization since activity could profoundly affect the rheology. In this letter, we report the realization of a 'real' 3-dimensional active glass composed of motile bacteria (*E. coli* strain RP4979). We measured the fluctuation and rheology of motile (active) and immotile (inactive) conditions, aiming to elucidate the way non-equilibrium activity influences the mechanics of the dense suspensions.

The *E. coli* strain RP4979 swims straight without tumbling [17,18], whereas flagellar motors are paralyzed in SHU321 [19]. Apart from flagellar rotation, other observable characteristics, such as the shape and size, are similar between them. Hereafter, we will refer to the motile and immotile bacterial suspensions as RP4979 and SHU321, respectively. To maintain the motility and viability of bacteria in dense suspensions, nutrients and metabolic by-products were continuously exchanged across the dialysis membrane as shown in Fig. 1a, S1a and b, and Supplementary. An optical-trapping force was applied to a colloidal particle ($2a$ = 3 μm diameter) dispersed in a bacterial suspension, and the motion of the probe particle was measured using a quadrant photodiode (QPD) (Fig. S1c). Feedback control of a sample stage was introduced to stably track a probe particle in a fluctuating environment [20,21].

We applied three types of microrheology (MR) [22-24]: passive MR, active MR, and force-clamp MR, as shown in Fig. 1a. In passive MR (PMR), the fluctuation of the probe velocity $v(t)$ was measured using a fixed probe laser ($\lambda$ = 830 nm), and the power spectral density (PSD) was calculated as $\langle |\tilde{v}(\omega)|^2 \rangle = \int_{-\infty}^{\infty} \langle v(0)v(t) \rangle e^{i\omega t} dt$ where ~ indicates Fourier transform [25]. In active MR (AMR), a sinusoidal force $F(t) = \hat{F}(\omega)\exp(-i\omega t)$ was applied to the probe particle using a drive laser ($\lambda$ = 1064 nm), and the velocity response $v(t) = \hat{v}(\omega)\exp(-i\omega t)$ was measured using the probe laser. Here, $\hat{F}(\omega)$ and $\hat{v}(\omega)$ are the complex oscillation amplitude at an angular frequency $\omega$ [24]. The complex mobility $R^*(\omega) = R'(\omega) - iR''(\omega) = \hat{v}(\omega)/\hat{F}(\omega)$ was then obtained. Hereafter, we refer to $\langle |\tilde{v}(\omega)|^2 \rangle$ and $R^*(\omega)$ as the fluctuation and response, respectively. In thermal equilibrium, they are related as $\langle |\tilde{v}(\omega)|^2 \rangle = 2k_B T R'(\omega)$ by the fluctuation-dissipation theorem (FDT). In force-clamp MR, a probe velocity $v$ was measured while applying a constant force $F$ to the probe with feedback-controlled optical trapping [26,27], as detailed in Supplementary. From the mobility $R(0) = v/F$, the viscosity $\eta$ was obtained by assuming Stokes' formula $\eta = F/6\pi av = 1/6\pi a R(0)$.

The velocity fluctuation $\langle|\tilde{v}(\omega)|^2\rangle$ and the response $R(\omega)$ are shown in Fig. 1b and 1c. For inactive SHU321 (Fig. 1b), the FDT $\langle|\tilde{v}(\omega)|^2\rangle = 2k_BTR'(\omega)$ was satisfied, indicating that the suspension was in equilibrium. In active RP4979 (Fig. 1c), the FDT was violated as $\langle|\tilde{v}(\omega)|^2\rangle > 2k_BTR'(\omega)$ at low frequencies, meaning that the suspension is driven out of equilibrium.

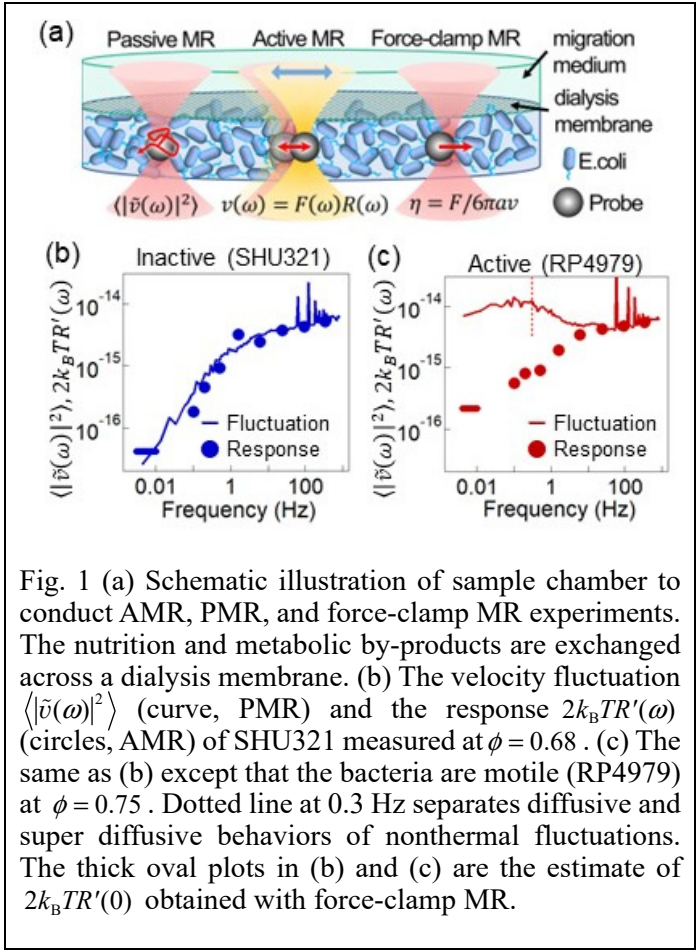

Fig. 1 (a) Schematic illustration of sample chamber to conduct AMR, PMR, and force-clamp MR experiments. The nutrition and metabolic by-products are exchanged across a dialysis membrane. (b) The velocity fluctuation $\langle|\tilde{v}(\omega)|^2\rangle$ (curve, PMR) and the response $2k_BTR'(\omega)$ (circles, AMR) of SHU321 measured at $\phi = 0.68$. (c) The same as (b) except that the bacteria are motile (RP4979) at $\phi = 0.75$. Dotted line at 0.3 Hz separates diffusive and super diffusive behaviors of nonthermal fluctuations. The thick oval plots in (b) and (c) are the estimate of $2k_BTR'(0)$ obtained with force-clamp MR.

The dynamics of bacterial suspensions were investigated by analyzing the microscope images as detailed in Supplementary. By calculating the Fourier transform of the pixel intensity $\tilde{I}(k,t)$, an intermediate scattering function $F(k,t) = \langle\tilde{I}(k,0)\tilde{I}^*(k,t)\rangle/\langle|\tilde{I}(k,0)|^2\rangle$ was obtained. Here, the wave number $k$ was chosen either corresponding to the probe-particle size ($k_1 = \pi/a$ μm$^{-1}$, Fig. 2a and 2b) or bacterial size ($k_0 = \pi/a_{bac}$ μm$^{-1}$, Fig. S2a and S2b) with $a_{bac} = 0.54$ μm. Assuming that the pixel intensity of an image correlates with the density of bacteria, it shows how the bacterial arrangement loses its memory. To investigate the dynamic heterogeneity, we computed the dynamic susceptibility $\chi_4(k_0,t)$ as shown in Fig. 2c and 2d [15,28-30].

The relaxation time of the bacterial fluctuation $\tau(k)$ was obtained by fitting the Kohlrausch-Williams-Watts function, $A\exp[-\{t/\tau(k)\}^B]$ to the measured $F(k,t)$. The dependence of $\tau(k_0)$ on $\phi$ ($\phi$: sample volume fraction relative to bacteria in a pellet) is shown in Fig. 2e. The colloidal glass transition is defined for inactive suspnsions when the relaxation time grows by $10^4 \sim 10^5$ times than that of a dilute condition and becomes too large to measure [3,31]. While $\tau(k_0)$ seems to increase exponentially for SHU321 like a strong glass former [32], $F(k_0,t)$ did not sufficiently relax within our experimental time scale (~ 100 s) for $\phi = 0.68, 0.92$.

Previous numerical studies predicetd that $\tau(k_0)$ in active suspensions remarkably differ from inactive suspensions only at high densities close to glass transition [4-13,33,34]. In experiments, $\tau(k_0)$ of active suspensions were remarkably smaller than that of inert ones even at lower fractions ($\phi = 0.35$ and 0.51, Fig. 2c). At these conditions, bacteria exhibited a turbulence-like collective motion [35,36] (Supplementary movies, Fig. S3). Such collective motion does not necessarily require local rearrangements of bacteria. Thus, $\tau(k_0)$ is underestimated as a structural relaxation time. At higher concentrations $\phi > 0.6$, turbulence-like motion was not observed in RP4979. $\tau(k_0)$ are still much smaller compared to that of SHU321 at the same $\phi$, indicating active fluidization [37-39].

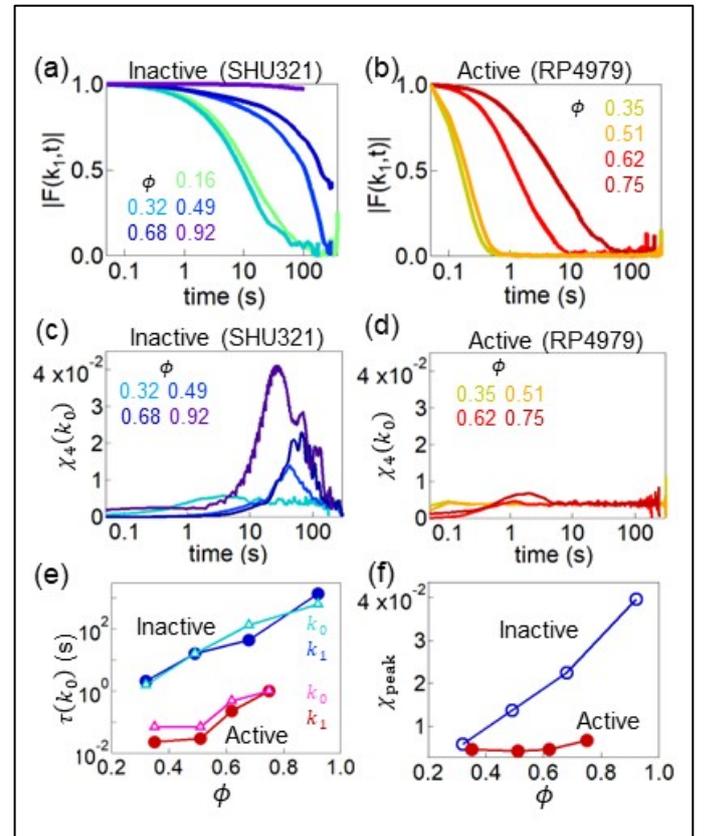

Fig. 2: Dynamics of bacterial suspensions investigated from microscope images. (a,b) Intermediate scattering function $F(k_1,t)$ in (a) inactive SHU321 and in (b) active RP4979 suspensions, respectively. The wave number $k_1$ corresponds to the probe particle's size. (c, d) Dynamic suceptibility $\chi_4(k_0,t)$, at the wave number $k_0$ corresponding to the bacterial size, measured in (c) inactive and (d) active suspensions. (e) Concentration ($\phi$) dependence of relaxation time $\tau(k_0)$ of bacterial fluctuation. $\tau(k_0)$ was obtained by fitting the KWW function $A\exp[-\{t/\tau(k)\}^B]$ to $F(k,t)$. Assuming $\tau(k) \propto k^{-2}$, $\tau(k_1)$ is also shown normalized as $\tau(k_1)(k_1/k_0)^2$. (f) Concentration ($\phi$) dependence of the peak height $\chi_{peak}$ of $\chi_4(k_0,t)$ which peaks around $t \sim \tau(k_0)$.

At high volume fractions ($\phi \gtrsim 0.5$), $\chi_4(k_0,t)$ peaks around $t \sim \tau(k_0)$ in SHU321 (Fig. 2c), and the peak height denoted by $\chi_{peak}$ grows, as shown in Fig. 2f. This behavior underscores the heterogeneous dynamics typical of glass formation. However, the peak was insignificant for RP4979 (Fig. 2d), indicating a more uniform dynamics in active suspensions [34,40]. Under the conditions exhibiting turbulence ($\phi = 0.35, 0.44, 0.51$), the susceptibility at the vortex length ($k_2 = \pi/10~\mu m^{-1}$) displayed pronounced collective dynamics, as shown in Fig. S2c.

The viscosity of SHU321 and RP4979 measured by the force-clamp MR is shown in Fig. 3. In (a), SHU321 at $\phi = 0.16$ and $0.32$ showed Newtonian viscosity in the range of forces, $0.1 \leq F < 10$ pN. At higher concentrations ($\phi = 0.49$ and $0.68$), $\eta$ decreased with increasing $F$. Similar non-Newtonian behavior, referred to as thinning or shear thinning, is typical of inactive glassy materials [41-43], but it disappeared in RP4979 which was Newtonian at all $\phi$ and $F$ (Fig. 3b). Notably, at large $F$, $\eta$ of concentrated SHU321 ($\phi \geq 0.49$) tends to converge to that of RP4979 at the same $\phi$ (Fig. S4). This tendency is reminiscent of the thinning of colloidal suspensions under external force application [44], despite the difference in how mechanical energy is injected into the system [45].

$\eta$ measured at small $F$ is shown in Fig. 3c (open circles: SHU321, closed circles: RP4979). The viscosity of SHU321 appears to depend exponentially on $\phi$, like a strong glass former. However, $\eta$ at high $\phi$ is underestimated since the suspensions are non-Newtonian. This suggests that the actual dynamics of SHU321 is more fragile. On the other hand, Newtonian $\eta$ of RP4979 increases only slowly with $\phi$, resulting in remarkable fluidization at high $\phi$. At $\phi \leq 0.4$, $\eta$ did not substantially differ between SHU321 and RP4979. In contrast to $\tau(k_0)$, the turbulence-like collective flows observed for $\phi = 0.35 \sim 0.51$ have marginal effects on the viscosity.

MR allows us to explore the fluctuation dynamics across a wide range of time and length scales [46]. In Fig. 1c, the fluctuation of RP4979 $\langle|\tilde{v}(\omega)|^2\rangle$ has negative slope above 0.3 Hz, indicating superdiffusion due to the unidirectional swimming of RP4979. A nearly flat plateau observed below 0.3 Hz suggests diffusion-like dynamics of non-thermal fluctuations, dominating $\langle|\tilde{v}(\omega)|^2\rangle$ at low frequencies. The diffusive dynamics of the non-thermal fluctuation appear due to the structural relaxation in dense suspensions. Nevertheless, at frequencies where the nonthermal fluctuation exhibits diffusive behavior, the RP4979 suspension is not fully fluidized, as evidenced by the positive gradient of the response $2k_B TR'(\omega)$ measured with AMR. As detailed in a prior study [47], this apparent inconsistency is ascribable to mesoscopic fluctuations triggered by structural relaxations. Similar mesoscopic fluctuations were also seen by PIV analysis of microscope image (Fig. S3).

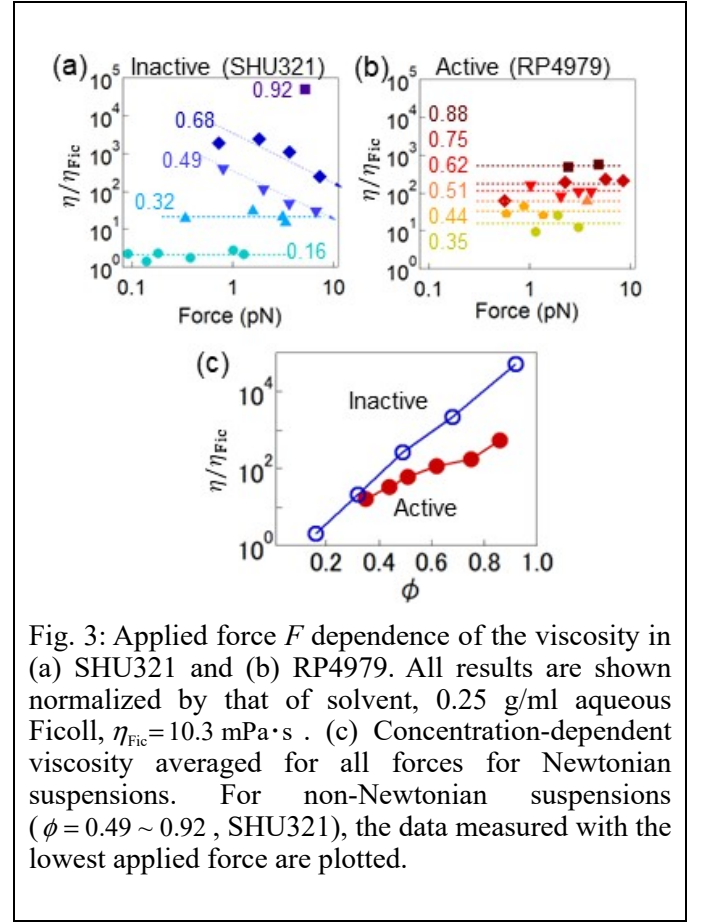

Fig. 3: Applied force $F$ dependence of the viscosity in (a) SHU321 and (b) RP4979. All results are shown normalized by that of solvent, 0.25 g/ml aqueous Ficoll, $\eta_{Fic} = 10.3$ mPa·s. (c) Concentration-dependent viscosity averaged for all forces for Newtonian suspensions. For non-Newtonian suspensions ($\phi = 0.49 \sim 0.92$, SHU321), the data measured with the lowest applied force are plotted.

Within a glassy environment, a structural relaxation gives rise to a long-range perturbation around the yield spot [48,49]. Whenever such a relaxation takes place within a specimen, it provokes a random displacement of a probe particle even if it is distant from the yield spot. This sequence of events imparts diffusive dynamics to the nonthermal fluctuations [50], at the time scale much smaller than that of structural relaxation [47]. Here, we determine "active" diffusion constant as $2D_A \equiv \lim_{\omega \to 0}\langle|\tilde{v}(\omega)|^2\rangle$ obtained from the plateau value of $\langle|\tilde{v}(\omega)|^2\rangle$. Extrapolation of $2k_B TR'(\omega)$ measured with AMR and $\lim_{\omega \to 0}\langle|\tilde{v}(\omega)|^2\rangle = 2k_B TR'(0)$ measured with particle pulling intersects at $\sim 0.01$ Hz, indicating that the typical frequency of alpha structural relaxation. At the time scale where the medium is fluidized, residual stress stored in a cage-like structure is released, and therefore the elastic resilience to the active stress is lost [47].

The wave-number dependence of $\tau(k)$ was obtained from microscope images. A dilute SHU321 suspension ($\phi = 0.16$) exhibits diffusive dynamics $\tau(k) = 1/D_s k^2$, as shown in Fig. 4a. Here, the self-diffusion constant $D_s$ of SHU321 was estimated as $D_s = k_B T/(6\pi\eta a_{bac})$ using $\eta$ measured with force-clamp MR. At higher concentration of SHU321, $\tau(k)$ becomes smaller than $1/D_s k^2$ while $\tau(k) \propto k^{-2}$ is still held (Fig. S5). This is typical behavior of non-dilute suspensions considering that $\tau(k)$ reflects collective diffusion [51,52]. In contrast, in a suspension with active turbulence ($\phi = 0.51$, RP4979), non-diffusive dependency $\tau(k) \propto k^{-1.34}$ was observed as shown in Fig. 4b, reflecting the collective dynamics in the suspension. At higher fractions ($\phi = 0.62, 0.75$), typical diffusive dynamics $\tau(k) \propto k^{-2}$ were recovered over the length scales greater than the bacteria. Furthermore, $\tau(k)$ agrees to $1/D'_A k^2$ (solid lines) where $D'_A = (a/a_{bac}) \cdot D_A$ is the active self-diffusion constant of RP4979 [53,54]. Interestingly, the nonthermal fluctuation in dense suspensions exhibits behavior similar to the thermal fluctuation in dilute suspensions although the underlying reason mechanism is elusive.

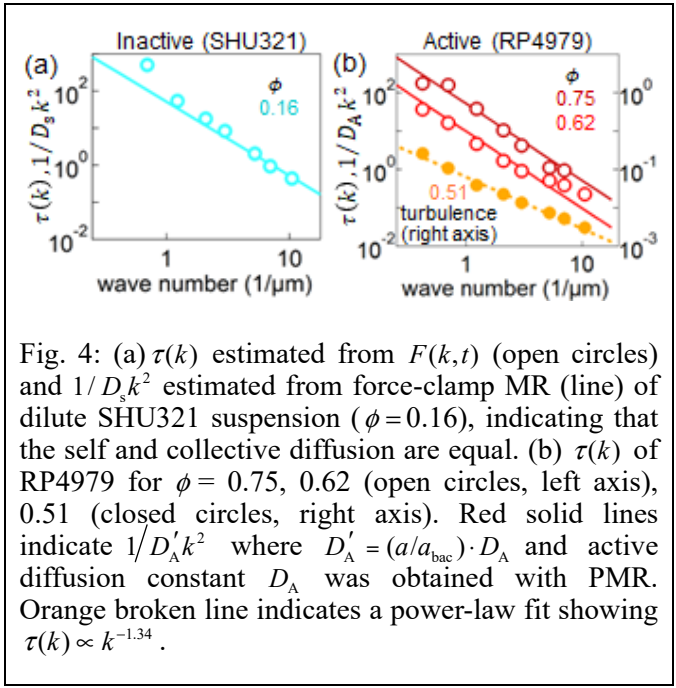

Fig. 4: (a) $\tau(k)$ estimated from $F(k,t)$ (open circles) and $1/D_s k^2$ estimated from force-clamp MR (line) of dilute SHU321 suspension ($\phi = 0.16$), indicating that the self and collective diffusion are equal. (b) $\tau(k)$ of RP4979 for $\phi = 0.75, 0.62$ (open circles, left axis), 0.51 (closed circles, right axis). Red solid lines indicate $1/D'_A k^2$ where $D'_A = (a/a_{bac}) \cdot D_A$ and active diffusion constant $D_A$ was obtained with PMR. Orange broken line indicates a power-law fit showing $\tau(k) \propto k^{-1.34}$.

Finally, we discuss how the active fluctuations impact the rheology of suspensions measured with AMR. In Fig. 5a and b, we present complex shear modulus $G^*(\omega) = G'(\omega) - iG''(\omega) = -i\omega/6\pi a R^*(\omega)$ measured with AMR. The range of frequencies where AMR was performed ($\gtrsim 0.1$ Hz) was above the typical frequency of the structural relaxation ($1/2\pi\tau_\alpha \sim 0.01$ Hz), as shown in Fig. 2a and b. SHU321 is stiffer than that of RP4979 in similar $\phi$, and showed smaller frequency dependency. Dense SHU321 is elastic at low frequencies as seen in Fig. 5c as $G'/G'' > 1$, which was not observed for RP4979 even at large $\phi$ (Fig. 5d). Instead, RP4979 is viscoelastic with a power-law contribution proportional to $G^*(\omega) \propto (-i\omega)^{1/2}$. The same power-law rheology was observed in dense suspensions driven far from equilibrium, e.g. cytoplasm in cells [20,39], dense star-like micelles under shear [55] and granular matter [56].

In theory, $G^*(\omega) \propto (-i\omega)^{1/2}$ was derived by analyzing the relaxation states of dense disordered materials at zero temperature [57]. At the critical jamming point, the relaxation times distribute indefinitely towards large time scales. Macroscopic shear deformations are given by the sum of these low-energy modes, which explains the power-law rheology. It was anticipated that the situation close to this setting was realized in dense suspensions of inactive particles such as emulsion, hydrogel and foam. However, in these materials, only $G''(\omega)$ follows the expected power-law but $G'(\omega)$ does not [58-60]. The elastic plateau appears in $G'(\omega)$ as we observed for SHU321 because thermal collisions between particles effectively increase the coordination number (effective contacts per particle) beyond the critical threshold for the jamming transition (twice the spatial dimension commonly termed the isostatic condition), imparting an entropic elasticity to the effective contacts [61].

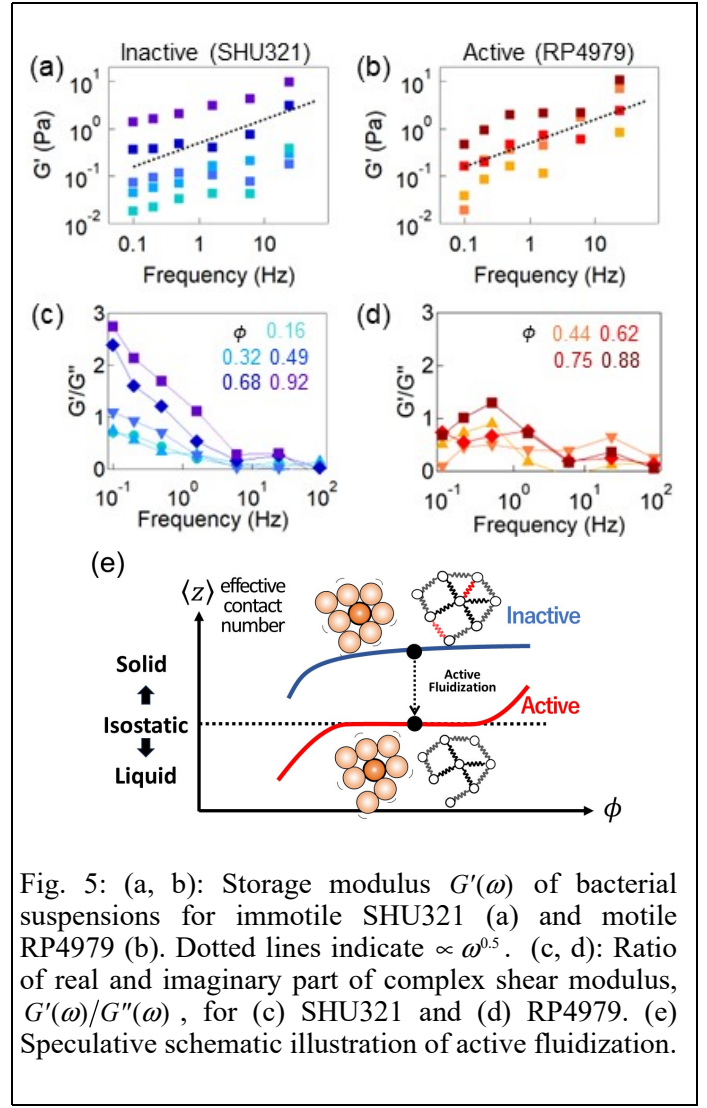

Fig. 5: (a, b): Storage modulus $G'(\omega)$ of bacterial suspensions for immotile SHU321 (a) and motile RP4979 (b). Dotted lines indicate $\propto \omega^{0.5}$. (c, d): Ratio of real and imaginary part of complex shear modulus, $G'(\omega)/G''(\omega)$, for (c) SHU321 and (d) RP4979. (e) Speculative schematic illustration of active fluidization.

On the other hand, nonthermal perturbations destabilize the effective contacts between bacteria, leading to yielding (Fig. 5e), while the homogeneous dynamics in active suspensions indicate a consistent coordination number. The mechanical energy generated by active objects (RP4979) is efficiently transferred to the mesoscopic fluctuations in the surrounding medium when the medium is solid [39,47,62,63]. Once fluidized, a decrease in energy-transfer efficiency halts further yielding, leaving the system close to the isostatic condition (Fig. 5e). Therefore, the nonthermal bacterial activity fluidizes the dense suspensions, such that they approach the transition boundary between solid (glass-like) and fluid states. The feedback mechanism between energy-transfer efficiency and the effective coordination number could autonomously adjust the system towards the emergence of self-organized criticality [64], as the observed power-law rheology suggests. While specifics warrant further investigations, the power-law rheology $G^*(\omega) \propto (-i\omega)^{1/2}$ is widely observed in dense active suspensions [39,55,56] likely because they share the similar non-equilibrium mechanism discussed as above.

In dense active suspensions near the pellet concentration, both fluctuations and rheological behavior were found to be influenced by activity. Consequently, the suspension shows a unique material property maintaining fluidity despite its high density, a property unattainable in inactive suspensions. A prominent example is the cytoplasm within living cells, which represents a densely packed suspension of active mechano-enzymes. Previous research has demonstrated that cytoplasm becomes more fluid with increasing metabolic activity [39]. In fact, the power-law rheology $G^*(\omega) \propto (-i\omega)^{1/2}$ has been observed in living cells with normal metabolic activity [20,39], while metabolically deficient cytoplasm exhibits a more elastic response akin to inactive dense suspensions [39]. The implications of a densely populated active fluid are profound, particularly in their potential to enhance cellular physiological processes. Through our MR experiments conducted in a simplified model system, we have gained valuable insights into the dynamic behavior of both colloidal and biological glasses from various perspectives of the data. Consequently, our experimental findings provided comprehensive support for the hypothesis that nonthermal mechanical energy plays a pivotal role in driving a dense suspension toward a delicate state, finely balanced at the boundary between fluid and solid.

**REFERENCES**

[1] S. Ramaswamy, in *Annual Review of Condensed Matter Physics, Vol 1*, edited by J. S. Langer2010), pp. 323.
[2] C. Bechinger, R. Di Leonardo, H. Lowen, C. Reichhardt, G. Volpe, and G. Volpe, Reviews of Modern Physics **88**, 45006 (2016).
[3] G. L. Hunter and E. R. Weeks, Reports on Progress in Physics **75**, 066501 (2012).
[4] L. Berthier and J. Kurchan, Nature Physics **9**, 310 (2013).
[5] R. Mandal, P. J. Bhuyan, M. Rao, and C. Dasgupta, Soft Matter **12**, 6268 (2016).
[6] R. Mandal, P. J. Bhuyan, P. Chaudhuri, M. Rao, and C. Dasgupta, Physical Review E **96**, 042605 (2017).
[7] S. K. Nandi, R. Mandal, P. J. Bhuyan, C. Dasgupta, M. D. Rao, and N. S. Gov, Proceedings of the National Academy of Sciences of the United States of America **115**, 7688 (2018).
[8] L. M. C. Janssen, Journal of Physics-Condensed Matter **31**, 503002 (2019).
[9] S. K. Nandi and N. S. Gov, Soft Matter **13**, 7609 (2017).
[10] A. Liluashvili, J. Onody, and T. Voigtmann, Physical Review E **96**, 062608 (2017).
[11] J. Reichert and T. Voigtmann, Soft Matter **17**, 10492 (2021).
[12] R. Mandal, P. J. Bhuyan, P. Chaudhuri, C. Dasgupta, and M. Rao, Nature Communications **11**, 2581 (2020).
[13] N. Klongvessa, F. Ginot, C. Ybert, C. Cottin-Bizonne, and M. Leocmach, Physical Review Letters **123**, 248004 (2019).
[14] N. Klongvessa, C. Ybert, C. Cottin-Bizonne, T. Kawasaki, and M. Leocmach, J Chem Phys **156**, 154509 (2022).
[15] E. Flenner, G. Szamel, and L. Berthier, Soft Matter **12**, 7136 (2016).
[16] R. Mandal and P. Sollich, Phys Rev Lett **125**, 218001 (2020).
[17] A. J. Wolfe, M. P. Conley, T. J. Kramer, and H. C. Berg, Journal of Bacteriology **169**, 1878 (1987).
[18] D. T. N. Chen, A. W. C. Lau, L. A. Hough, M. F. Islam, M. Goulian, T. C. Lubensky, and A. G. Yodh, Physical Review Letters **99**, 148302 (2007).
[19] K. A. Datsenko and B. L. Wanner, Proc Natl Acad Sci U S A **97**, 6640 (2000).
[20] K. Nishizawa, M. Bremerich, H. Ayade, C. F. Schmidt, T. Ariga, and D. Mizuno, Sci Adv **3**, e1700318 (2017).
[21] Y. Sugino, M. Ikenaga, and D. Mizuno, Appl Sci-Basel **10**, 4970 (2020).
[22] F. Gittes and C. F. Schmidt, Opt Lett **23**, 7 (1998).
[23] T. G. Mason, K. Ganesan, J. H. vanZanten, D. Wirtz, and S. C. Kuo, Physical Review Letters **79**, 3282 (1997).
[24] D. Mizuno, D. A. Head, F. C. MacKintosh, and C. F. Schmidt, Macromolecules **41**, 7194 (2008).
[25] B. Schnurr, F. Gittes, F. C. MacKintosh, and C. F. Schmidt, Macromolecules **30**, 7781 (1997).
[26] N. Honda, K. Shiraki, F. Van Esterik, S. Inokuchi, H. Ebata, and D. Mizuno, New J Phys **24**, 053031 (2022).
[27] K. Nishizawa, N. Honda, S. Inokuchi, H. Ebata, T. Ariga, and D. Mizuno, Physical Review E **108**, 034601 (2023).
[28] S. C. Glotzer, V. N. Novikov, and T. B. Schroder, Journal of Chemical Physics **112**, 509 (2000).
[29] A. S. Keys, A. R. Abate, S. C. Glotzer, and D. J. Durian, Nature Physics **3**, 260 (2007).
[30] T. Narumi, S. V. Franklin, K. W. Desmond, M. Tokuyama, and E. R. Weeks, Soft Matter **7**, 1472 (2011).
[31] J. Mattsson, H. M. Wyss, A. Fernandez-Nieves, K. Miyazaki, Z. B. Hu, D. R. Reichman, and D. A. Weitz, Nature **462**, 83 (2009).
[32] C. A. Angell, Science **267**, 1924 (1995).
[33] Y. Koyano, H. Kitahata, and A. S. Mikhailov, Epl **128**, 40003 (2019).
[34] N. Oyama, T. Kawasaki, H. Mizuno, and A. Ikeda, Physical Review Research **1**, 032038(R) (2019).
[35] H. H. Wensink, J. Dunkel, S. Heidenreich, K. Drescher, R. E. Goldstein, H. Lowen, and J. M. Yeomans, Proceedings of the National Academy of Sciences of the United States of America **109**, 14308 (2012).
[36] J. Dunkel, S. Heidenreich, K. Drescher, H. H. Wensink, M. Bar, and R. E. Goldstein, Physical Review Letters **110**, 228102 (2013).
[37] B. R. Parry, I. V. Surovtsev, M. T. Cabeen, C. S. O'Hem, E. R. Dufresne, and C. Jacobs-Wagner, Cell **156**, 183 (2014).
[38] K. Nishizawa, K. Fujiwara, M. Ikenaga, N. Nakajo, M. Yanagisawa, and D. Mizuno, Sci Rep-Uk **7**, 15143 (2017).

[39] H. Ebata, K. Umeda, K. Nishizawa, W. Nagao, S. Inokuchi, Y. Sugino, T. Miyamoto, and D. Mizuno, Biophys J **122**, 1781 (2023).
[40] E. Flenner, G. Szamel, and L. Berthier, Soft Matter **12**, 7136 (2016).
[41] L. G. Wilson, A. W. Harrison, W. C. K. Poon, and A. M. Puertas, Epl **93**, 58007 (2011).
[42] D. Winter, J. Horbach, P. Virnau, and K. Binder, Physical Review Letters **108**, 028303 (2012).
[43] N. J. Wagner and J. F. Brady, Physics Today **62**, 27 (2009).
[44] R. N. Zia, Annu Rev Fluid Mech **50**, 371 (2018).
[45] P. K. Morse, S. Roy, E. Agoritsas, E. Stanifer, E. I. Corwin, and M. L. Manning, Proc Natl Acad Sci U S A **118** (2021).
[46] H. Seyforth, M. Gomez, W. B. Rogers, J. L. Ross, and W. W. Ahmed, Physical Review Research **4**, 023043 (2022).
[47] K. Umeda, K. Nishizawa, W. Nagao, S. Inokuchi, Y. Sugino, H. Ebata, and D. Mizuno, Biophys J **122**, 4395 (2023).
[48] C. E. Maloney and A. Lemaitre, Phys Rev E Stat Nonlin Soft Matter Phys **74**, 016118 (2006).
[49] A. Lemaitre and C. Caroli, Phys Rev Lett **103**, 065501 (2009).
[50] K. Martens, L. Bocquet, and J. L. Barrat, Phys Rev Lett **106**, 156001 (2011).
[51] A. J. Banchio, G. Nagele, and J. Bergenholtz, Journal of Chemical Physics **113**, 3381 (2000).
[52] C. W. J. Beenakker and P. Mazur, Physica A **126**, 349 (1984).
[53] A. J. Levine and T. C. Lubensky, Physical Review Letters **85**, 1774 (2000).
[54] D. A. Head and D. Mizuno, Physical Review E **81**, 041910 (2010).
[55] A. R. Jacob, A. S. Poulos, A. N. Semenov, J. Vermant, and G. Petekidis, J Rheol **63**, 641 (2019).
[56] P. Marchal, N. Smirani, and L. Choplin, J Rheol **53**, 1 (2009).
[57] B. P. Tighe, Phys Rev Lett **107**, 158303 (2011).
[58] A. J. Liu, S. Ramaswamy, T. G. Mason, H. Gang, and D. A. Weitz, Phys Rev Lett **76**, 3017 (1996).
[59] K. Krishan, A. Helal, R. Hohler, and S. Cohen-Addad, Phys Rev E Stat Nonlin Soft Matter Phys **82**, 011405 (2010).
[60] H. M. Wyss, K. Miyazaki, J. Mattsson, Z. Hu, D. R. Reichman, and D. A. Weitz, Phys Rev Lett **98**, 238303 (2007).
[61] L. J. Wang and N. Xu, Soft Matter **9**, 2475 (2013).
[62] L. M. C. Janssen, J Phys Condens Matter **31**, 503002 (2019).
[63] T. Ariga, M. Tomishige, and D. Mizuno, Phys Rev Lett **121**, 218101 (2018).
[64] D. Sornette, *Critical phenomena in natural sciences : chaos, fractals, selforganization, and disorder : concepts and tools* (Springer, Berlin ; New York, 2004), 2nd edn., Springer series in synergetics,.